\documentclass{article}
\usepackage{amsmath}
\usepackage{graphicx}
\usepackage{hyperref}
\usepackage{subfig}

\begin{document}

\title{The Kernighan-Lin Search Algorithm\footnote{This paper is a
  slightly edited version of an AI class project report submitted at
  the University of Illinois at Urbana-Champaign on May 12, 1995.}}

\author{Ali Dasdan\\
  KD Consulting\\
  Saratoga, CA, USA\\
  alidasdan@gmail.com\\
}

\maketitle

\begin{abstract}
The traveling salesman problem (TSP) and the graph partitioning
problem (GPP) are two important combinatorial optimization problems
with many applications. Due to the NP-hardness of these problems,
heuristic algorithms are commonly used to find good, or hopefully
near-optimal, solutions. Kernighan and Lin have proposed two of the
most successful heuristic algorithms for these problems: The
Lin-Kernighan (LK) algorithm for TSP and the Kernighan-Lin (KL)
algorithm for GPP. Although these algorithms are problem specific to
TSP and GPP, they share a problem-agnostic mechanism, called variable
depth search, that has wide applicability for general search. This
paper expresses this mechanism as part of a general search algorithm,
called the Kernighan-Lin Search algorithm, to facilitate its use
beyond the TSP and GPP problems. Experimental comparisons with other
general search algorithms, namely, genetic algorithms, hill climbing,
and simulated annealing, on function optimization test suites confirm
that the new algorithm is very successful in solution quality and
running time.
\end{abstract}

\section{Introduction}

The traveling salesman problem (TSP) and the graph partitioning
problem (GPP) are two important combinatorial optimization
problems. These problems are encountered in many situations, for
example, the first phase in VLSI physical design involves partitioning
of the circuit elements into some number of blocks for subsequent
phases; this partitioning phase yields a substantial reduction in the
complexity of the VLSI layout problem. These problems are also served
as test-beds to evaluate the performance of new (heuristic)
algorithms.

Since TSP and GPP are NP-hard problems, heuristic algorithms are used
to obtain good, or hopefully near-optimal, solutions. Brian Kernighan
and Shen Lin, who worked at the famed Bell Labs at the time, proposed
a heuristic algorithm for each problem. The algorithm for TSP has been
referred to as the Lin-Kernighan (LK) algorithm~\cite{LiKe73}, and the
algorithm for GPP has been referred to as the Kernighan-Lin (KL)
algorithm~\cite{KeLi70}. These algorithms are iterative improvement
algorithms (also called local search algorithms) in that they
iteratively improve a given initial solution in search of a locally
optimal solution.

There are a large number of other approaches (sometimes called
``metaheuristics''~\cite{Wi20}) to solve these problems such as
genetic algorithms~\cite{Go89}, simulated
annealing~\cite{JoArMc89,JoArMc91}, mean-field annealing, tabu search,
and integer programming, e.g., see
\cite{LoMaSt02,PiRo10,SoSeGl18,Wi20}. However, the LK and KL
algorithms have remained to be among the best approaches when solution
quality and running time are considered together.

A big part of the success of the LK and KL algorithms seems to be a
mechanism called ``variable depth search'', which usually lead to
better solutions than those possible with a vanilla hill climbing
algorithm. In this paper, I present this mechanism as a general local
search algorithm and apply to function optimization problems to
experimentally prove the generality of the approach and facilitate its
use beyond TSP and GPP. I will refer to this algorithm as the
Kernighan-Lin Search (KLS) algorithm.

This paper also presents the results of experiments done on a standard
test suite of seven test cases (popular in the genetic algorithms
community in 1990s) in comparison to other optimization algorithms:
Genetic algorithms (GAs), simulated annealing (SA), and hill climbing
(or hill-climbing). I also implemented and tested two new versions of
the KLS algorithm by incorporating problem-specific knowledge into the
KLS algorithm.

The experimental results support the initial intuition about the
performance of the KLS algorithm. The KLS algorithm performed very
well on all the test cases. When its solution quality and running time
are considered together, it usually outperforms the simulated
annealing algorithm.

The rest of the paper is organized as follows. A very short overview
the related work and how this paper came about in \S~\ref{sec:related}
and \S~\ref{sec:historical}. The fundamentals of the local search
strategy and my knowledge representation for the experiments are
introduced in \S~\ref{sec:local} and \S~\ref{sec:know-rep}. These
sections are the basis of the following four sections on the specific
algorithms: Hill Climbing in \S~\ref{sec:hill-algo}, the proposed
Kernighan-Lin search algorithms in \S~\ref{sec:kls-algo}, simulated
annealing in \S~\ref{sec:sa-algo}, and genetic algorithms in
\S~\ref{sec:gen-algo}. These sections are followed by the experimental
framework in \S~\ref{sec:exper} and the experimental results and
discussion in \S~\ref{sec:results}. The paper ends with the
conclusions and future work in \S~\ref{sec:conclusions}.

\section{Related Work\protect\footnote{This section was originally written in
  1995 but got revised with new references in 2025.}}
\label{sec:related}

The technical literature on metaheuristics is vast, e.g., see
\cite{AhErOr02,LoMaSt02,MaPaRe18,PiRo10,SoSeGl18,Wi20} for
overviews.

Variable depth search was proposed by Kernighan and Lin and used in
\cite{KeLi70} for graph partitioning and in \cite{LiKe73} for
traveling salesman problem. In the latter reference, the variable
depth search is introduced as ``a general approach to heuristics that
we believe to be of wide applicability,'' and its ``considerable
success'' with graph partitioning, citing the former reference, is
also celebrated. It is interesting that it took almost two decades for
variable depth search mechanism to be used for problems other than
these two.

The complexity of finding good solutions using local search
techniques, including the KL and LK algorithms, is explored in
\cite{JoPaYa88,PaScYa90} under a new complexity class called
``polynomial local search'' (PLS).

Early references recognizing the benefits of variable depth search
include \cite{AhErOr02} and \cite{PiRo10} as a part of very-large
neighborhood search techniques, \cite{LoMaSt01} as part of the
iterated local search metaheuristic, \cite{ThOr89} as a special case
of the cyclic transfers metaheuristic, \cite{Gl96} as a special case
of the ejection chains metaheuristic, \cite{Gr03} and \cite{MeFr02} to
improve the performance of genetic algorithms.

Papers related to this work include the short survey of applications
in \cite{AhErOr02}, the applications in \cite{Gr03} and \cite{MeFr02}
to improve the performance of genetic algorithms, and the applications
to the vehicle routing problems in \cite{VaLeSc90,VaLeSc93}, the
quadratic assignment problem in \cite{Sk90} and more recently in
\cite{GoGo12,MeFr02,Pa09}, the maximum clique problem in
\cite{KaSaNa07}, and nurse rostering in \cite{BuCuQu07}.

Given this paper was originally published in 1995, it may also be
considered as an early reference for the variable depth search
mechanism and its general applications.

For further study, interested readers should consult these references
together with a huge number of references available on Arxiv, Google
Scholar, and generic search engines. A search query containing the
phrase ``variable depth search'' can help filter irrelevant results.

\section{Historical Note}
\label{sec:historical}

My own experience with metaheuristics was due to two studies I
completed during my masters graduate work in early 1990s: New
algorithms for multiway graph and hypergraph
partitioning~\cite{DaAy97} and genetic synthesis of unsupervised
learning algorithms~\cite{DaOf93a,DaOf93b}. The code for the former
was open sourced in \cite{Da18a,Da18b}. This experience involved the
Kernighan-Lin algorithm and its variants; search and optimization
algorithms such as local search, simulated annealing, genetic
algorithms, neural networks; and combinatorial optimization
problems. 

I originally submitted this paper as a class project report for the
artificial intelligence class taught by Professor Gerald
DeJong~\footnote{Prof. DeJong is one of the pioneers of machine
learning with the explanation-based learning approach.} at the
University of Illinois at Urbana-Champaign on May 12,
1995. Unfortunately I did not get around to submitting a paper for
publication and had lost the paper copy of the report for many
years. Recently I discovered the lost paper copy and immediately
scanned it into the pdf format. This paper is a slightly updated
version of the original class project report. Since I do not have the
software code used in this work anymore, I was not able to replicate
the experiments on the current computers.

It is been 30 years since this paper was originally written. It is my
hope that the main part of this paper, namely, the KLS algorithm, may
serve as an early reference (from 1995) for iterative improvement
algorithms using variable depth search.

\section{Local Search Preliminaries}
\label{sec:local}

A combinatorial optimization (minimization) problem can be specified
by identifying a set of {\em solution}s $S$ and a {\em cost function}
$f(S)$ that assigns a {\em cost} to each solution $S$. The goal is to
find a solution of the {\em optimal cost} $S^{*}$, i.e., a solution
with the minimum possible cost for minimization problems. There may be
more than one such solution.

Many combinatorial optimization problems are hard in that it takes a
great amount of time, usually exponential in the size of the problem,
to find a solution to even very small problem instances. As a result,
we resort to heuristic algorithms to find good solutions that are
hopefully close to the optimal solution. These heuristic algorithms
(or heuristics) usually run in a low-order polynomial time.

The methods used for designing heuristics tend to be rather problem
specific. {\em Local search} is one of the few general heuristics. It
is usually based on trial-and-error. Local search is the basis of the
Kernighan-Lin Search (KLS) algorithm and the other algorithms
discussed in this paper.

The first choice that must be made in order to apply a local search
algorithm to a certain problem is the choice of a knowledge
representation in which the problem and its solutions are encoded in
some form, as will be explained in the next section.

The second choice is to define a suitable {\em neighborhood structure}
on the knowledge representation; this structure specifies a
neighborhood of solutions around each given solution. These neighbor
solutions are in some sense close to that given solution, and each can
be obtained by perturbing the given solution in one step on the
knowledge representation. Each such perturbation is called a {\em
  move}. If the perturbation leads to a solution with a higher cost,
then the move is an {\em uphill move}; otherwise, it is a {\em
  downhill move}. A solution $S$ in a neighborhood structure is {\em
  locally optimal} (locally minimal in this work) if none of its
neighbors has a lower cost. If a locally optimal solution is the best
solution to the problem, it is called the {\em (globally) optimal
  solution}, which is very hard to find in most cases given the
NP-hardness of the problem.

The third choice for a local search algorithm is the choice of
devising an algorithm to generate an initial solution to the
problem. This algorithm must be a low-order polynomial time algorithm.
Initial solutions are almost always generated randomly though such
initial solutions can have a very bad cost.

Starting from a given initial solution $S$, a local search algorithm
tries to find a better solution in $N(S)$, the neighborhood of $S$. If
a better neighbor is found, a search starts for a better neighbor of
that one, and so on. Since the set of solutions is finite, this search
must halt, that is, the algorithm must come up with a locally optimal
solution. Since the local search algorithms iteratively improve an
initial solution in search of a locally optimal one, they are also
called {\em iterative improvement algorithms}.

A neighbor solution $S'$ in the neighborhood $N(S)$ of a solution $S$
can be found in three different ways:
\begin{enumerate}
\item the first descent method: $S'$ is the first solution in $N(S)$ that has
a lower cost than that of $S$,

\item the {\em steepest descent} method: all the solutions in $N(S)$
  are examined and $S'$ is the solution with the best (lowest) cost,
  and

\item $S'$ is a solution randomly chosen among those in $N(S)$.
\end{enumerate}

The hill climbing algorithm and the KLS algorithm use variations of
the steepest descent method, and the simulated annealing and genetic
algorithms use variations of the third method. Obviously, the steepest
descent method takes more time than the other two. The main difference
between the hill climbing algorithm and the KLS algorithm is that the
former will choose the first lower cost solution while the latter will
traverse a path of moves over all the solutions in the neighborhood,
even uphill moves, and then choose the solution on the path from the
first solution that leads to the maximum cost reduction (called the
maximum gain). The reason for calling the KLS method ``variable
depth'' is because it is not known in advance which solution on the
path will lead to the maximum gain; in other words, the depth of the
chosen solution from the starting solution on the path is variable per
neighborhood.

\section{Knowledge Representation}
\label{sec:know-rep}

Since I implemented the hill climbing algorithm, the KLS algorithm,
and the simulated annealing algorithm myself, I used the same
representation for all of them. Since I used a third-party genetic
algorithms software package for the genetic algorithm, I had to adapt
the representation imposed by this package but the representation was
the same. Note that the representation of solutions is also called the
{\em encoding} of the solutions.

Consider a function $f(x_1, x_2, \cdots, x_n)$ of $n$ variables, where
each variable is also called a {\em coefficient}. Each coefficient is
a real number. The function $f$ gives the cost of a solution that is
obtained by using these $n$ coefficients. These coefficients can also
be considered as the parameters of the problem. I encoded each
coefficient using $l$ bits where $l$ depends on the problem at
hand. Thus, the code that I used for $f$ is simply a concatenation of
the individual codes for the coefficients, i.e, the code for $f$ is a
linear bit string of length $nl$. I assume that each solution thus
encoded is a feasible solution to the problem, which holds for all the
test functions I experimented with. Since each bit can take two
values, the code for $f$ can encode $2^{nl}$ solutions, which gives
the size of the search space.

Each coefficient thus coded is an integer in $[0,2^{l}-1]$. To map each
  coefficient $x_{i}$ into a real number in $[L, U]$, where $L$ is the
  lower bound and $U$ is the upper bound on every coefficient, I used
  the following linear mapping:
\begin{equation}
x_{i} = L + int(x_{i})\frac{(U-L)}{(2^{l}-1)}
\end{equation}
where $int(x_{i})$ is the integer value of the $l$ bits encoding the
coefficient $x_{i}$. This is also the mapping used in
GAucsd~\cite{ScGr92}.

Let $S$ be the code for $f(x_1, x_2, \cdots, x_n)$ as described
above. This code encodes a solution for $f$. Also let $S(j)$ denote
the $j$th bit of $S$. Then, we can perturb $S$ to generate the
neighbor solutions in $N(S)$ by flipping a single bit of $S$ in each
step. Each bit flipping defines a move, and each move generates a
solution in the neighborhood. A neighbor solution $S'$ obtained by
flipping the $j$th bit of $S$ has the $j$th bit value of $S'(j) =
1-S(j)$. Since there are $nl$ bits in $S$, there are $nl$ moves and
correspondingly $nl$ neighbor solutions in $N(S)$.

Note that in reality there are $2^{nl}$ settings of $nl$ bits. This
obviously defines the entire search space and it is too large, i.e.,
exponential, to search in the neighborhood of a given
solution. Actually searching the entire search space is what amounts
to an exhaustive search, and it invalidates the purpose of using any
of the algorithms in this paper. The goal for effective local search
is to define a neighborhood that is small enough for efficiency and
large enough for coverage. This is the case for our neighborhood
definition in that since we are using $nl$ bit flips as the moves, our
neighborhood size is reduced from exponential to polynomial.

The cost difference $\Delta f = f(S)-f(S')$ that results from the
$j$th move (the $j$th bit flip) is called the {\em gain} of that move
and denoted by $g$. The gain of a move can be negative or positive. A
positive gain means that the move produced a better solution, and a
negative gain means that the move produced a worse solution. Thus, a
move with positive gain is a downhill move, and a move with negative
gain is an uphill move.

\begin{figure}[t]
  \centering
  \includegraphics[scale=0.4]{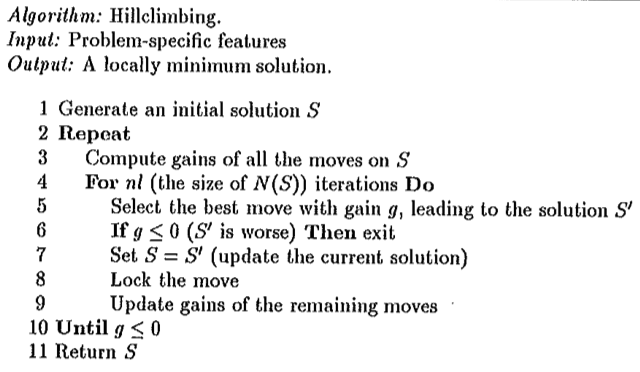}
  \caption{The hill climbing algorithm.}
  \label{fig:hill-algo}
\end{figure}

\section{Hill Climbing}
\label{sec:hill-algo}

Hill climbing or the hill climbing algorithm corresponds exactly to
the local search strategy with the steepest descent method. Since all
the problems in this work are minimization problems, the hill climbing
algorithm here is actually a hill ``descent'' algorithm, akin to
gradient descent. The hill climbing algorithm used in this work is
given in Fig.~\ref{fig:hill-algo}.

In step 1 in Fig.~\ref{fig:hill-algo}, an initial solution $S$ is
generated randomly. Each bit of $S$ is set to $0$ or $1$ with equal
probability. This initialization scheme was also employed for the
other algorithms.

Step 8 of the algorithm, locking the move, is necessary to test every
solution in the neighborhood at most once in each iteration of the
repeat loop. After every iteration of the for loop in step 4, the
number of possible solutions to test decreases by one. The hill
climbing algorithm only allows downhill moves. This often leads it to
getting stuck in poor local optima.

\begin{figure}[t]
  \centering
  \includegraphics[scale=0.4]{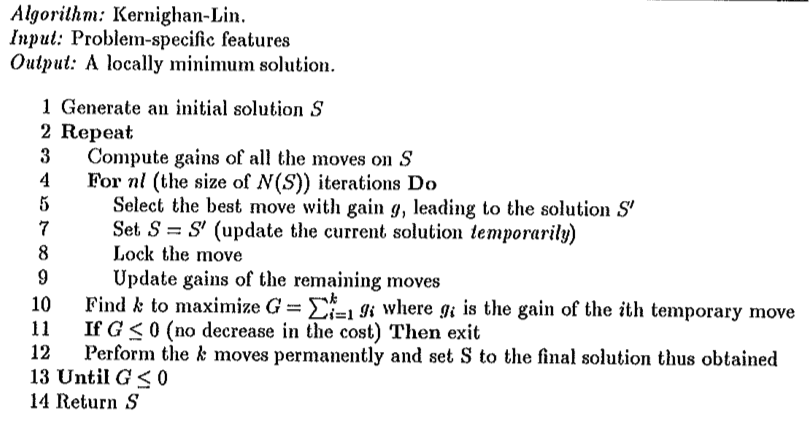}
  \caption{The Kernighan-Lin search algorithm.}
  \label{fig:kls-algo}
\end{figure}

\section{Kernighan-Lin Search (KLS)}
\label{sec:kls-algo}

As discussed earlier in this paper, the Kernighan-Lin Search (KLS)
algorithm is a generalization of the variable depth search mechanism
of the Kernighan-Lin and Lin-Kernighan algorithms. The KLS algorithm
is given in Fig.~\ref{fig:kls-algo}. The steps of the KLS algorithm
match those of the hill climbing algorithm except for the steps 7, 10,
and 12 in the former.

The key point or mechanism in the KLS algorithm is as follows: In step
7, the solution update is done {\em temporarily} since we do not know
how many moves are necessary to get the best solution starting from
the initial solution S. This number of moves is obtained in step 10 by
finding the maximum prefix sum of the gain sequence. The $k$ moves
yielding the maximum gain sum in step 10 are made permanent in step
12.

Note that since $k$ is not known in advance, it is the {\em variable}
in ``variable depth search'' and it can be at most $nl$, the total
number of moves available in the neighborhood structure starting from
a given solution in the neighborhood. The KLS algorithm, unlike the
hill climbing algorithm, does not stop when it gets a negative gain in
step 5; this negative gain may be included in the maximum prefix
computed in step 10, Thus, the KLS algorithm sometimes employs uphill
moves, which seems to be a key reason behind its success.

Step 8, locking the move, is to prevent cycling, i.e., visiting the
same moves again, during neighborhood search. This mechanism is also
present in the hill climbing algorithm. As shown in \cite{DaAy97},
locking can actually be relaxed or removed altogether to yield even
better solutions. In this paper, we will not discuss this extension
further.

To incorporate problem-specific knowledge into the KLS algorithm, I
also designed two additional versions of the KLS algorithm. The first
version (called the KLS1 algorithm) is based on the fact that the
coefficients optimizing each function are integers. The definition of
a move is different in the KLS1 case. A move can either increase or
decrease a coefficient by $1$ provided that the value of the
coefficient is still between the upper and lower bounds. If both moves
are possible, the one yielding higher gain is selected.

The second version (called the KLS2 algorithm) is based on the fact
that all the test functions in my test suite are symmetric in that the
coefficients optimizing them are equal. Thus, instead of applying
moves to each coefficient, I applied moves only the first coefficient
and copied the first coefficient to the others. The moves were bit
flips as in the KLS algorithm.

Both of these versions led to a drastic reduction in the size of the
search space for each function. However, their performances differed
greatly, as will be shown in the experimental results section.

\begin{figure}[t]
  \centering
  \includegraphics[scale=0.4]{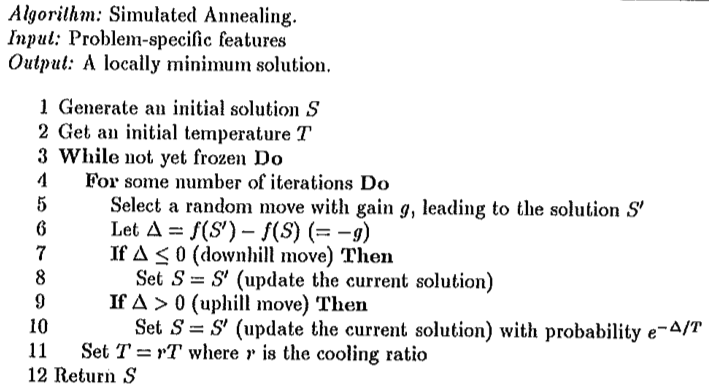}
  \caption{The simulated annealing algorithm.}
  \label{fig:sa-algo}
\end{figure}

\section{Simulated Annealing (SA)}
\label{sec:sa-algo}

The simulated annealing (SA) algorithm~\cite{JoArMc89,JoArMc91} is
also a local search algorithm, but it allows occasional uphill moves
to avoid getting trapped in poor local optima. The uphill moves are
determined by both the gain of the current move and a control
parameter called temperature. The simulated annealing algorithm used
in this work is given in Fig.~\ref{fig:sa-algo}.

Although not fully shown in Fig.~\ref{fig:sa-algo}, the simulated
annealing algorithm actually contains nearly a dozen parameters
together with the temperature.  The tests in both step 3 and step 4
are determined using some of these parameters. I used the simulated
annealing algorithm and the parameter setting suggested in
\cite{JoArMc89,JoArMc91}. For an explanation of these parameters,
refer to the same references.

\begin{table}[ht]
\centering
\begin{tabular}{r|l}
\hline
{\bf Parameter} & {\bf Setting}\\
\hline\hline
Structure Length & problem specific\\
\hline
Total Trials & 60000 (about 2000 generations)\\
\hline
Population Size & 30\\
\hline
Crossover Rate & 0.850\\
\hline
Mutation Rate & 0.005\\
\hline
Generation Gap & 1.000\\
\hline
Max. Generations without Evaluation & 2\\
\hline
Maximum Bias & 0.990\\
\hline
Convergence Threshold & 0.950\\
\hline
Scaling Window & -1\\
\hline
Sigma Scaling & 1\\
\hline
DPE Time Constant & 0\\
\hline
Max Convergence & the same as Structure Length\\
\hline
Options & Abelcu\\
\hline
Coding & Gray coding (not binary)\\
\end{tabular}
\caption{Parameter settings for GAucsd.}
\label{tab:ga-algo}
\end{table}

\section{Genetic Algorithm (GA)}
\label{sec:gen-algo}

Genetic Algorithms (GAs)~\cite{Go89} are search and optimization
algorithms working by mimicking the process of natural evolution as a
means of advancing toward the optimum. They can search large and
complex spaces effectively. They are robust in that they adapt to a
wide variety of environments. 

To implement the genetic algorithm in this work, I used the GAucsd
software package~\cite{ScGr92}, which provides a genetic algorithm for
function minimization. This package in particular and genetic
algorithms in general have lots of parameters or knobs to tune for
good performance. Since \cite{ScCaEs89} provides robust settings after
extensive experiments on the same test suite as mine, I used the
settings suggested there. For reference, the list of the parameters
and their settings is also given in Table~\ref{tab:ga-algo}. The
meaning of each parameter is explained in \cite{ScGr92}.

\section{Experimental Framework}
\label{sec:exper}

Except for the genetic algorithm, I implemented the hill climbing,
KLS, and simulated annealing algorithms all in the C programming
language. All the experiments were carried out on a SUN SPARCstation
(a high end workstation computer in early 1990s).  I performed tests
on seven functions. The first five functions (F1 through F5) are
called De Jong’s test suite~\cite{Ac89,De75}. This test suite was a
standard test suite in the genetic algorithms community in 1990s.  The
last two functions (F6 and F7) are from \cite{ScCaEs89}. These two
functions were specifically designed to optimize the control
parameters of genetic search. The task for all the functions is the
minimization. The properties of these functions are listed below:
\begin{enumerate}

\begin{figure}[ht]
  \centering
  \subfloat[F1]{{\includegraphics[scale=0.15]{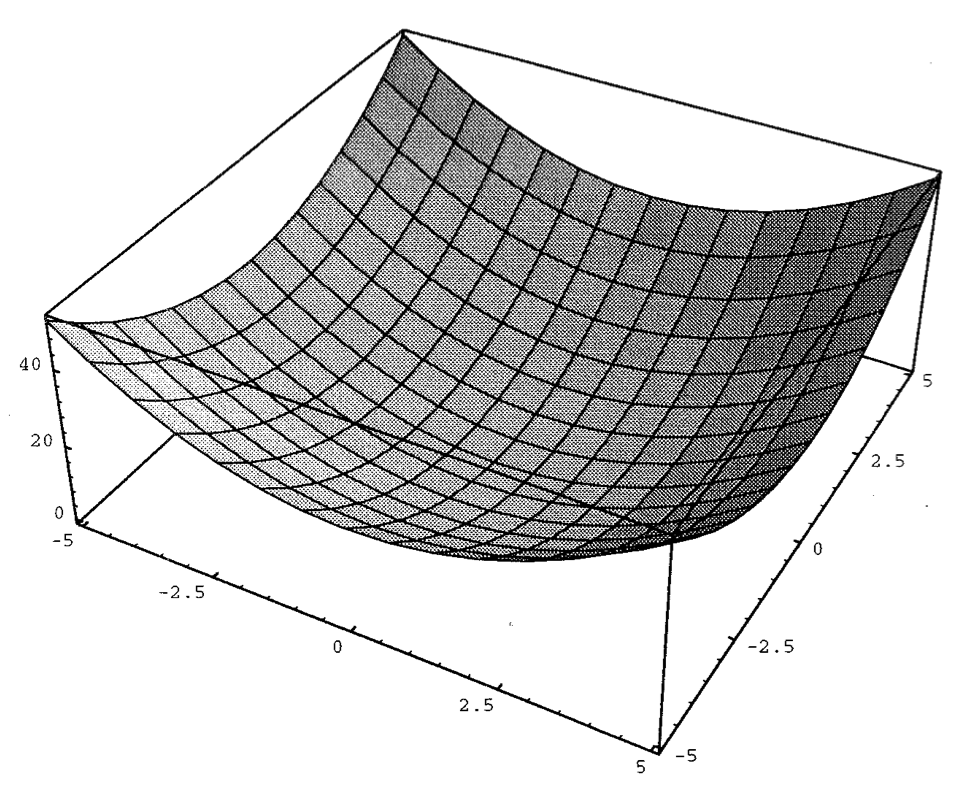}}}
  \qquad
  \subfloat[F2]{{\includegraphics[scale=0.15]{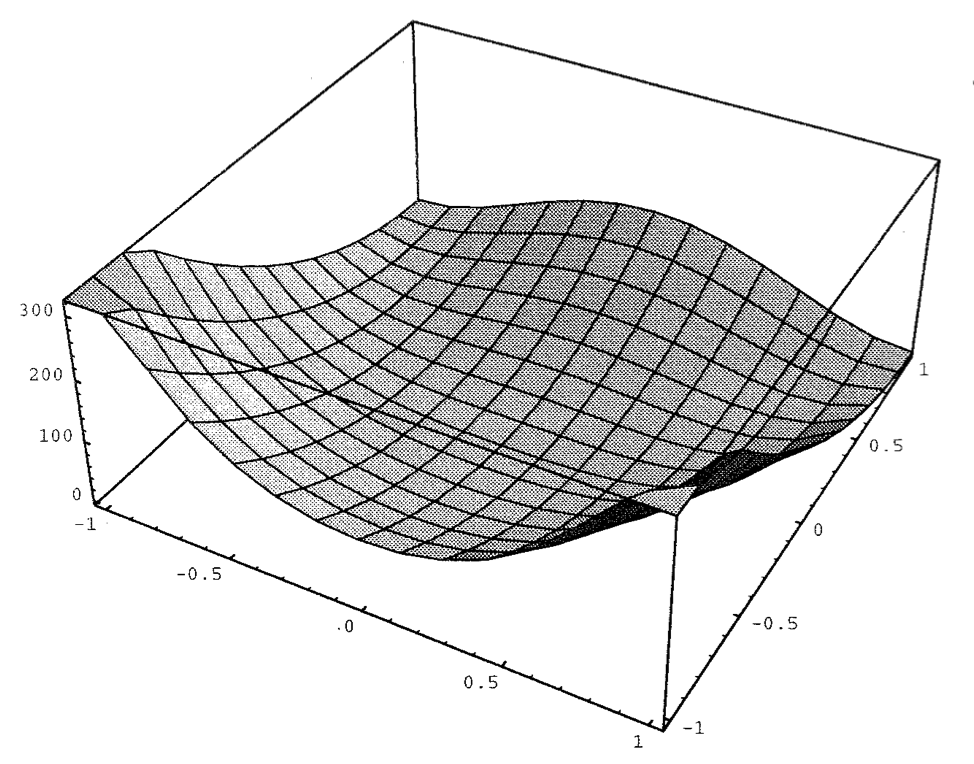}}}  
  \caption{Three-dimensional illustrations of the functions F1 and
    F2 in our test suite.}
  \label{fig:f1f2-3d}
\end{figure}

\item F1 is an instance of the “parabola”, as illustrated in
  Fig.~\ref{fig:f1f2-3d}(a). It is a continuous, convex, unimodal,
  quadratic, low-dimensional, and deterministic function. Its minimum
  value is zero at the origin. It contains three coefficients and each
  coefficient is coded in $10$ bits. This function is defined as
\begin{equation}
f_{1}(x)=\sum_{i=1}^{3}x_{i}^{2}
\end{equation}
where $-5.12\leq x_{i}\leq 5.12$ for each $i$.

\item F2 is an instance of “Rosenbrock's saddle”, as illustrated in
  Fig.~\ref{fig:f1f2-3d}(b). It is a continuous, non-convex, unimodal,
  quartic, low-dimensional, and deterministic function. Its minimum
  value is zero at $(1,1)$. It contains two coefficients and each
  coefficient is coded in $12$ bits. This function is defined as
\begin{equation}
f_{2}(x)=100(x_{1}^{2}-x_{2})^2+(1-x_{1})^2
\end{equation}
where $-2.048\leq x_{i}\leq 2.048$ for each $i$.

\begin{figure}[ht]
  \centering
  \subfloat[F3]{{\includegraphics[scale=0.15]{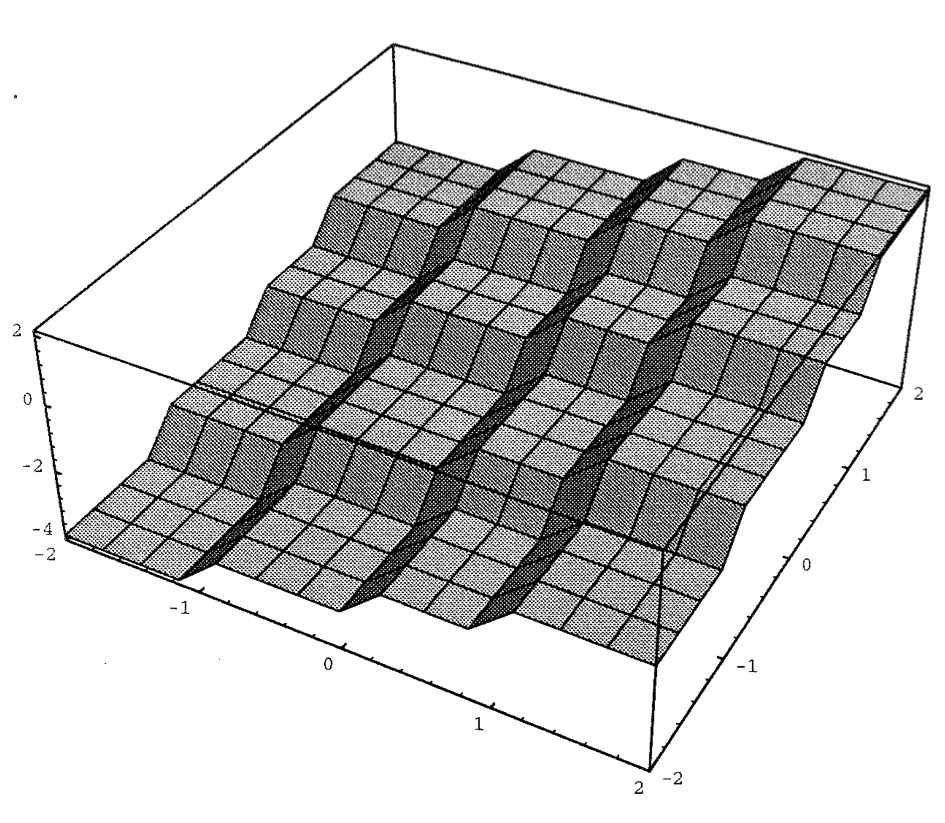}}}
  \qquad
  \subfloat[F4 (without noise)]{{\includegraphics[scale=0.15]{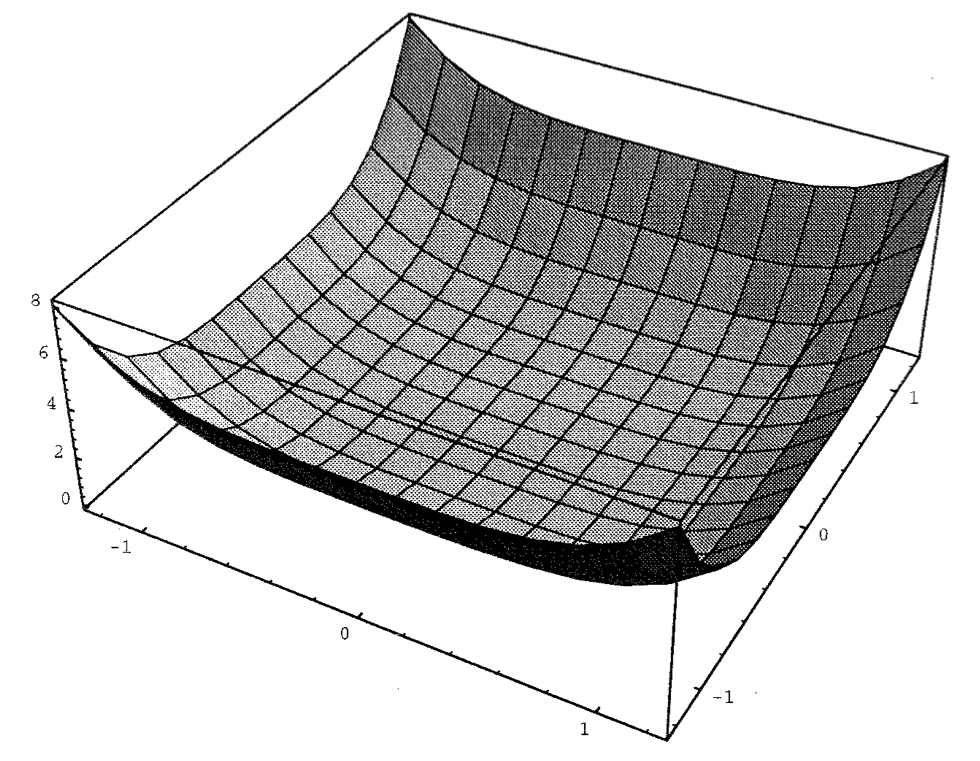}}}
  \caption{Three-dimensional illustrations of the functions F3 and
    F4 in our test suite.}
  \label{fig:f3f4-3d}
\end{figure}

\item F3 is an instance of the “step function”, as illustrated in
  Fig.~\ref{fig:f3f4-3d}(a). It is a discontinuous, non-convex,
  unimodal, low-dimensional, and deterministic function. Its minimum
  value is $-30$ but I added $30$ to the output to move the minimum
  value to $0$. It contains five coefficients and each coefficient is
  coded in $10$ bits. This function is defined as
\begin{equation}
f_{3}(x)=\sum_{i=1}^{5}\lfloor x_{i}\rfloor
\end{equation}
where $-5.12\leq x_{i}\leq 5.12$ for each $i$. In my implementation

\item F4 is an instance of the “quartic with noise”, as illustrated in
  Fig.~\ref{fig:f3f4-3d}(b). It is a continuous, convex, unimodal,
  quartic, high-dimensional, and stochastic function (with Gaussian
  noise). Its expected minimum value is zero but it may go as low as
  $-4$, depending on the Gaussian noise generated, i.e., four standard
  deviations away from the zero mean in the negative direction. It
  contains $30$ coefficients and each coefficient is coded in $8$
  bits. This function is defined as
\begin{equation}
f_{4}(x)=\sum_{i=1}^{30}ix_{i}^4+Gauss(0,1)
\end{equation}
where $-1.28\leq x_{i}\leq 1.28$ for each $i$, and $Gauss(0, 1)$ is a
random number drawn from the standard normal distribution with a mean
of $0$ and a standard deviation of $1$.

\begin{figure}[ht]
  \centering
  \includegraphics[scale=0.15]{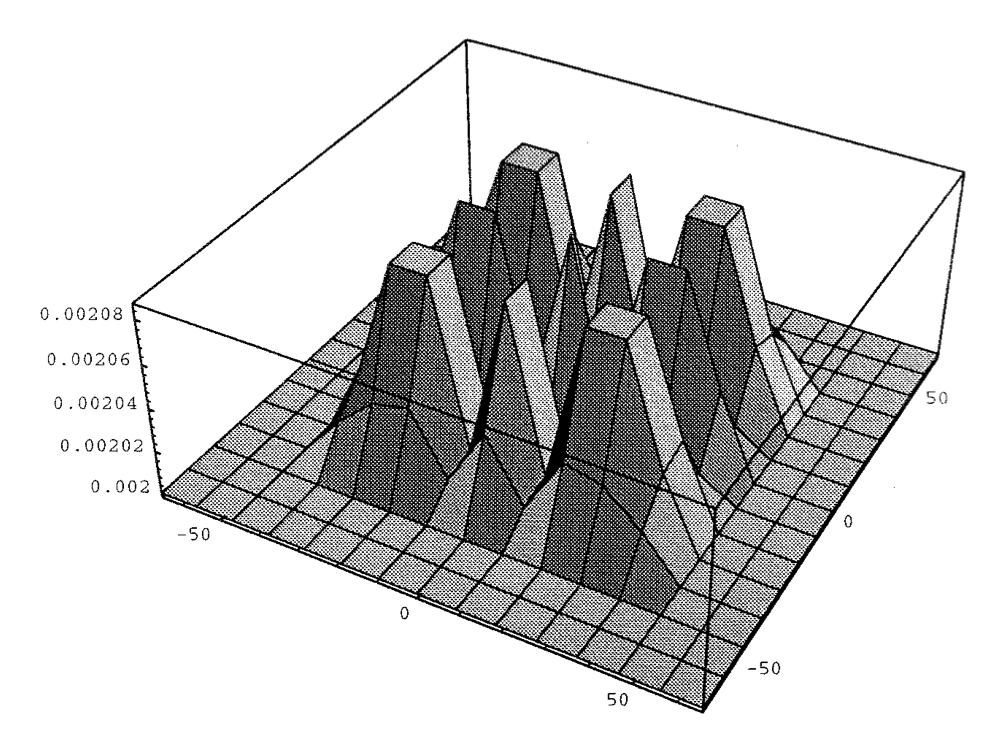}
  \caption{Three-dimensional illustration of the function F5 in our
    test suite.}
  \label{fig:f5-3d}
\end{figure}

\item F5 is an instance of “Shekel’s foxholes”, as illustrated in
  Fig.~\ref{fig:f5-3d}. It is a continuous, non-convex, multimodal,
  non-quadratic, high-dimensional, and deterministic function. Its
  minimum value is $1$. It contains 25 coefficients and each
  coefficient is coded in $17$ bits. This function is defined as
\begin{equation}
f_{5}(x)=(\frac{1}{500}+\sum_{j=1}^{25}(j+\sum_{i=1}^2(x_{i}-a_{ij})^6)^{-1})^{-1}
\end{equation}
where $-65.536\leq x_{i}\leq 65.536$ for each $i$, and $a$ is an array
of integers, where $a_{1j}=(-32,-16,0,16,32,-32,-16,\cdots,0,16,32)$
and $a_{2j}=(-32,-32,-32,-32,-32,-16,-16,\cdots,32,32,32)$.

\begin{figure}[ht]
  \centering
  \subfloat[3d]{{\includegraphics[scale=0.15]{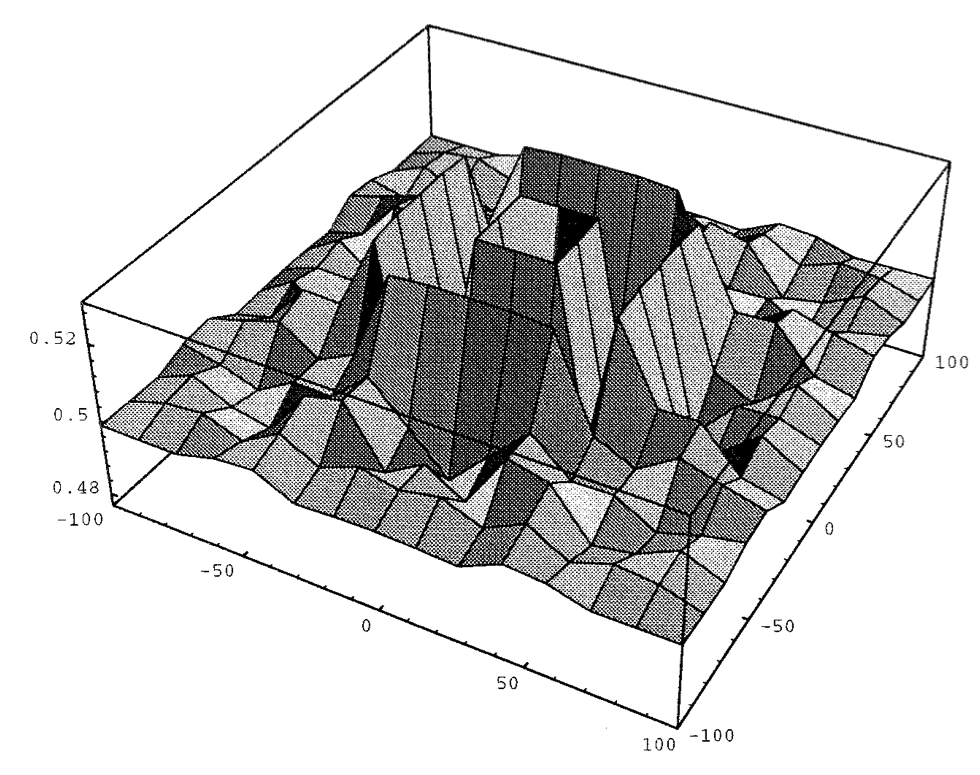}}}
  \qquad
  \subfloat[2d]{{\includegraphics[scale=0.15]{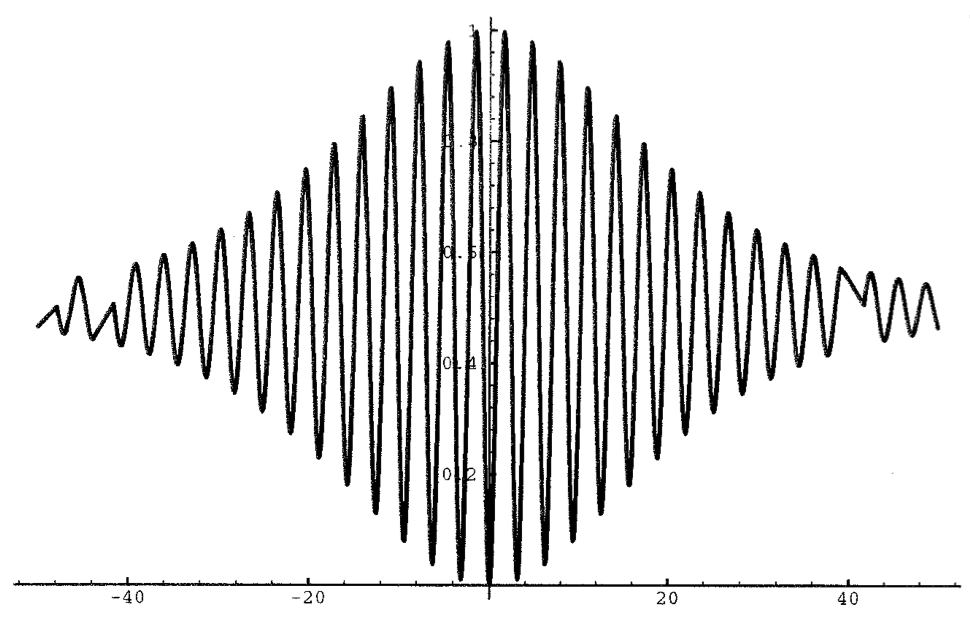}}}  
  \caption{Three- and two-dimensional illustrations of the function F6
    in our test suite.}
  \label{fig:f6-3d}
\end{figure}

\item F6 is an instance of the “sine envelope sine wave”, as
  illustrated in three- and two-dimensions in Fig.~\ref{fig:f6-3d}(a)
  and (b), respectively. It is a continuous, non-convex, multimodal,
  low-dimensional, and deterministic function. Its minimum value is
  zero. It contains two coefficients and each coefficient is coded in
  $22$ bits. This function is defined as
\begin{equation}
f_{6}(x)=0.5+\frac{sin^{2}(\sqrt{x_{1}^{2}+x_{2}^{2}})-0.5}{(1.0+0.001(x_{1}^{2}+x_{2}^{2}))^{2}}
\end{equation}
where $-100.0\leq x_{i}\leq 100.0$ for each $i$.

\begin{figure}[ht]
  \centering
  \subfloat[3d]{{\includegraphics[scale=0.15]{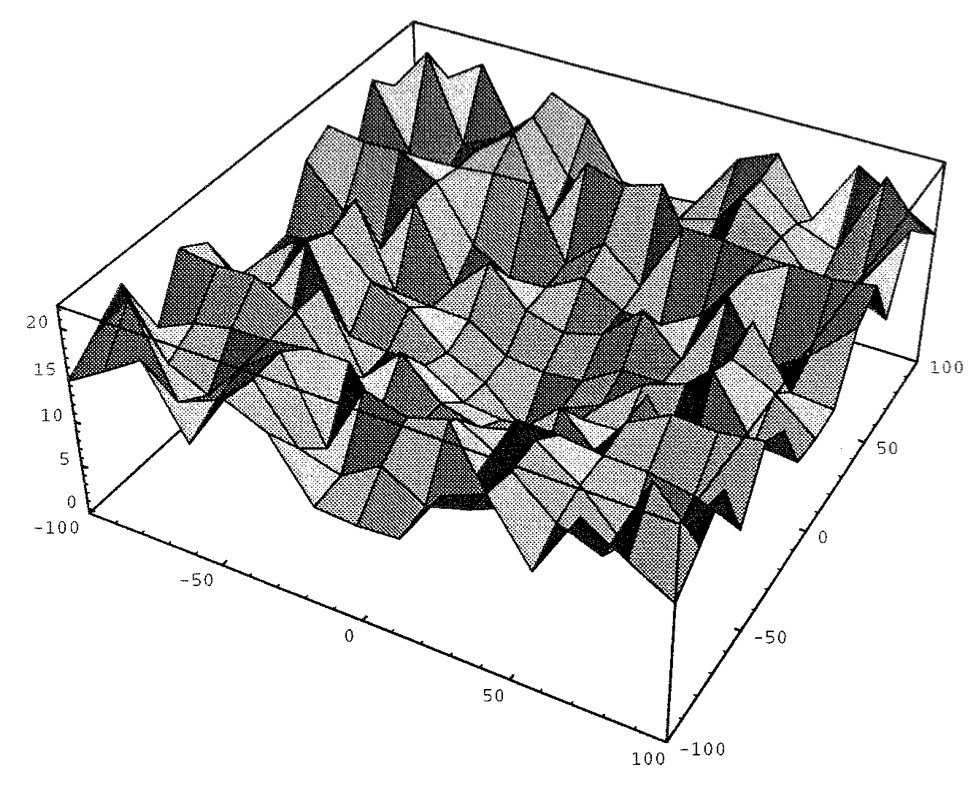}}}
  \qquad
  \subfloat[2d]{{\includegraphics[scale=0.15]{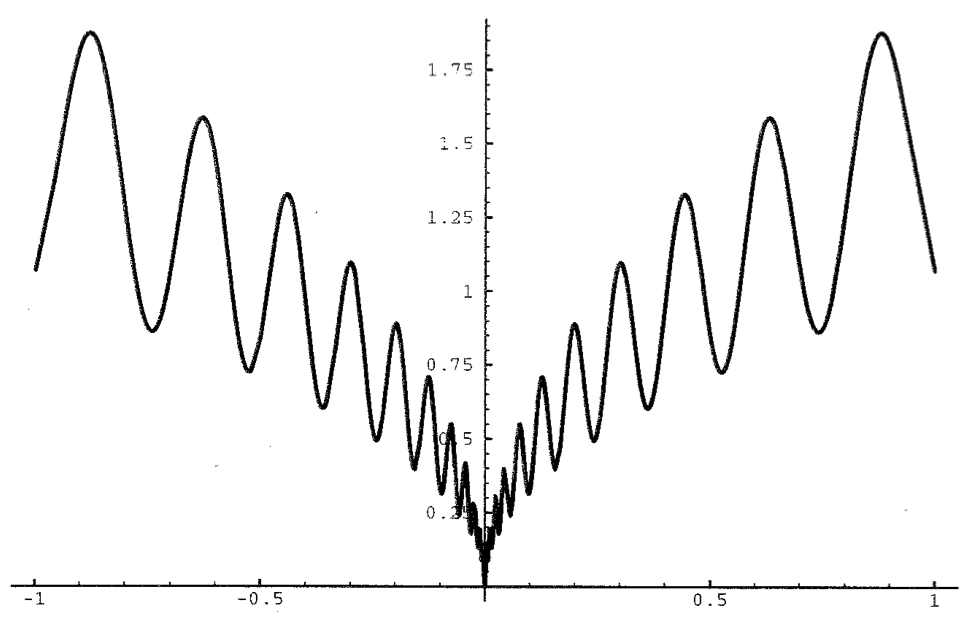}}}  
  \caption{Three- and two-dimensional illustrations of the function F7
    in our test suite.}
  \label{fig:f7-3d}
\end{figure}

\item F7 is an instance of the “stretched V sine wave”, as illustrated
  in three- and two-dimensions in Fig.~\ref{fig:f7-3d}(a) and (b),
  respectively. It is a continuous, non-convex, multimodal,
  low-dimensional, and deterministic function. Its minimum value is
  zero. It contains two coefficients and each coefficient is coded in
  $22$ bits. This function is defined as
\begin{equation}
f_{7}(x)=(x_{1}^{2}+x_{2}^{2})^{0.25}(sin^{2}(50(x_{1}^{2}+x_{2}^{2})^{0.1})+1.0)
\end{equation}
where $-100.0\leq x_{i}\leq 100.0$ for each $i$.
\end{enumerate}

For all the algorithms (including the genetic algorithm) compared in
this work, the same coding and function implementations were
used. Each algorithm was run 50 times on each function, starting from
different random seeds. The running time of each algorithm was
measured by the UNIX \verb|time| command.

A note on the search space size reduction using the KLS1 and KLS2
versions of the KLS algorithm follows: The size of the search space
for F1 in the KLS algorithm is $2^{30}$. In the case of the KLS1
algorithm, the size reduces to about $11^{3}\approx 2^{10}$, and in
the case of the KLS2 algorithm, the size also reduces to $2^{10}$. The
reductions are more significant some of the other test functions.

\begin{table}[ht]
\centering
\begin{tabular}{r|rrrr|rr}
\hline
\multicolumn{7}{c}{{\bf Results on F1}}\\
\hline\hline
\multicolumn{1}{c}{} &
\multicolumn{4}{|c|}{Cost} &
\multicolumn{2}{c}{Time (s)}\\
\multicolumn{1}{c|}{Algorithm}
& \multicolumn{1}{c}{Avr}
& \multicolumn{1}{c}{Std}
& \multicolumn{1}{c}{Min}
& \multicolumn{1}{c}{Max}
& \multicolumn{1}{|c}{Avr}
& \multicolumn{1}{c}{Std} \\
\hline
KLS2 & 0.00 & 0.00 & 0.00 & 0.00 &  0.00 & 0.00\\
SA   & 0.00 & 0.00 & 0.00 & 0.00 & 13.26 & 0.63\\
GA   & 0.00 & 0.00 & 0.00 & 0.00 &  5.54 & -\\
KLS  & 0.00 & 0.00 & 0.00 & 0.00 &  0.00 & 0.00\\
KLS1 & 0.00 & 0.00 & 0.00 & 0.00 &  0.00 & 0.00\\
Hill & 0.00 & 0.00 & 0.00 & 0.00 &  0.00 & 0.00\\
\hline

\multicolumn{7}{c}{{\bf Results on F2}}\\
\hline\hline
\multicolumn{1}{c}{} &
\multicolumn{4}{|c|}{Cost} &
\multicolumn{2}{c}{Time (s)}\\
\multicolumn{1}{c|}{Algorithm}
& \multicolumn{1}{c}{Avr}
& \multicolumn{1}{c}{Std}
& \multicolumn{1}{c}{Min}
& \multicolumn{1}{c}{Max}
& \multicolumn{1}{|c}{Avr}
& \multicolumn{1}{c}{Std} \\
\hline
KLS2 & 0.00 &  0.00 & 0.00 &   0.00 &  0.00 & 0.00\\
GA   & 0.01 &  0.01 & 0.00 &   0.04 &  8.58 & -\\
KLS  & 0.16 &  0.28 & 0.00 &   0.79 &  0.00 & 0.00\\
SA   & 0.45 &  0.45 & 0.00 &   1.00 & 11.08 & 1.73\\
Hill & 1.58 &  2.18 & 0.00 &  10.04 &  0.00 & 0.00\\
KLS1 & 8.64 & 27.87 & 0.00 & 109.00 &  0.00 & 0.00\\
\hline

\end{tabular}
\caption{Solution quality (cost) and running time (time) for each
  algorithm to minimize the functions F1 and F2 with the minimum
  values of $0$ and $0$, respectively. For both functions, all the
  algorithms hit the minimum values. For F1, the average solution
  quality of each algorithm is the same. For both cases, the
  algorithms are ranked from the best to the worst on the average
  solution quality.}
\label{tab:f1f2}
\end{table}

\begin{table}[ht]
\centering
\begin{tabular}{r|rrrr|rr}
\hline
\multicolumn{7}{c}{{\bf Results on F3}}\\
\hline\hline
\multicolumn{1}{c}{} &
\multicolumn{4}{|c|}{Cost} &
\multicolumn{2}{c}{Time (s)}\\
\multicolumn{1}{c|}{Algorithm}
& \multicolumn{1}{c}{Avr}
& \multicolumn{1}{c}{Std}
& \multicolumn{1}{c}{Min}
& \multicolumn{1}{c}{Max}
& \multicolumn{1}{|c}{Avr}
& \multicolumn{1}{c}{Std} \\
\hline
GA   &  0.00 & 0.00 & 0.00 &  0.00 & 15.16 & -\\
SA   &  5.00 & 0.00 & 5.00 &  5.00 &  7.28 & 0.45\\
KLS  &  5.08 & 0.27 & 5.00 &  6.00 &  0.00 & 0.00\\
KLS2 &  5.90 & 1.92 & 5.00 & 10.00 &  0.00 & 0.00\\
Hill &  7.54 & 0.92 & 6.00 &  9.00 &  0.00 & 0.00\\
KLS1 & 16.10 & 8.23 & 0.00 & 31.00 &  0.00 & 0.00\\
\hline

\multicolumn{7}{c}{{\bf Results on F4}}\\
\hline\hline
\multicolumn{1}{c}{} &
\multicolumn{4}{|c|}{Cost} &
\multicolumn{2}{c}{Time (s)}\\
\multicolumn{1}{c|}{Algorithm}
& \multicolumn{1}{c}{Avr}
& \multicolumn{1}{c}{Std}
& \multicolumn{1}{c}{Min}
& \multicolumn{1}{c}{Max}
& \multicolumn{1}{|c}{Avr}
& \multicolumn{1}{c}{Std} \\
\hline
KLS1 & -0.41 & 0.57 & -1.57 &  0.84 &   1.40 &  0.66\\
KLS2 & -1.88 & 0.43 & -2.66 & -1.03 &   0.00 &  0.00\\
SA   & -2.31 & 0.34 & -3.14 & -1.69 & 114.88 &  4.60\\
KLS  & -2.56 & 0.35 & -3.37 & -1.59 &  33.44 & 11.71\\
GA   & -2.96 & 0.23 & -3.45 & -2.56 &  40.66 & -     \\
Hill &  4.21 & 2.94 & -0.88 & 15.58 &   0.00 &  0.00\\
\hline

\end{tabular}
\caption{Solution quality (cost) and running time (time) for each
  algorithm to minimize the functions F3 and F4 with the minimum value
  of $0$ (after my change to the original F3) and the expected minimum
  value of $0$, respectively. For both cases, the algorithms are
  ranked from the best to the worst on the average solution
  quality. Due to the statistical nature of F4, the solution quality
  is measured as the distance from the expected minimum value of
  zero.}
\label{tab:f3f4}
\end{table}

\begin{table}[ht]
\centering
\begin{tabular}{r|rrrr|rr}
\hline
\multicolumn{7}{c}{{\bf Results on F5}}\\
\hline\hline
\multicolumn{1}{c}{} &
\multicolumn{4}{|c|}{Cost} &
\multicolumn{2}{c}{Time (s)}\\
\multicolumn{1}{c|}{Algorithm}
& \multicolumn{1}{c}{Avr}
& \multicolumn{1}{c}{Std}
& \multicolumn{1}{c}{Min}
& \multicolumn{1}{c}{Max}
& \multicolumn{1}{|c}{Avr}
& \multicolumn{1}{c}{Std} \\
\hline
GA   & 1.00 & 0.00 & 1.00 &  1.00 &  25.80 & -    \\
KLS2 & 1.93 & 3.17 & 1.00 & 12.67 &   0.00 &  0.00\\
KLS  & 2.98 & 3.13 & 1.00 & 11.72 & 107.04 & 50.67\\
KLS1 & 4.37 & 3.68 & 1.00 & 12.67 &   1.50 &  1.27\\
Hill & 6.62 & 5.70 & 1.00 & 18.30 &   1.14 &  0.35\\
SA   & -    & -    & -    & -     & -      & -    \\
\hline

\multicolumn{7}{c}{{\bf Results on F6}}\\
\hline\hline
\multicolumn{1}{c}{} &
\multicolumn{4}{|c|}{Cost} &
\multicolumn{2}{c}{Time (s)}\\
\multicolumn{1}{c|}{Algorithm}
& \multicolumn{1}{c}{Avr}
& \multicolumn{1}{c}{Std}
& \multicolumn{1}{c}{Min}
& \multicolumn{1}{c}{Max}
& \multicolumn{1}{|c}{Avr}
& \multicolumn{1}{c}{Std} \\
\hline
SA   & 0.01 & 0.00 & 0.00 & 0.01 & 65.32 & 6.04\\
KLS  & 0.02 & 0.03 & 0.00 & 0.13 &  0.00 & 0.00\\
KLS2 & 0.07 & 0.11 & 0.00 & 0.31 &  0.00 & 0.00\\
Hill & 0.20 & 0.15 & 0.00 & 0.47 &  0.00 & 0.00\\
KLS1 & 0.47 & 0.05 & 0.31 & 0.50 &  0.00 & 0.00\\
GA   & -    & -    & -    & -    & -     & -   \\
\hline

\end{tabular}
\caption{Solution quality (cost) and running time (time) for each
  algorithm to minimize the functions F5 and F6 with minimum values of
  $1$ and $0$, respectively. For both cases, the algorithms are ranked
  from the best to the worst on the average solution quality.}
\label{tab:f5f6}
\end{table}

\begin{table}[ht]
\centering
\begin{tabular}{r|rrrr|rr}

\hline
\multicolumn{7}{c}{{\bf Results on F7}}\\
\hline\hline
\multicolumn{1}{c}{} &
\multicolumn{4}{|c|}{Cost} &
\multicolumn{2}{c}{Time (s)}\\
\multicolumn{1}{c|}{Algorithm}
& \multicolumn{1}{c}{Avr}
& \multicolumn{1}{c}{Std}
& \multicolumn{1}{c}{Min}
& \multicolumn{1}{c}{Max}
& \multicolumn{1}{|c}{Avr}
& \multicolumn{1}{c}{Std} \\
\hline
KLS2 & 0.00 & 0.00 & 0.00 &  0.00 &  0.00 & 0.00\\
SA   & 0.01 & 0.01 & 0.00 &  0.02 & 23.70 & 4.57\\
KLS  & 0.16 & 0.59 & 0.00 &  2.51 &  0.04 & 0.20\\
Hill & 3.38 & 3.51 & 0.01 & 10.65 &  0.00 & 0.00\\
KLS1 & 8.44 & 2.07 & 2.00 & 11.31 &  0.00 & 0.00\\
GA   & -    & -    & -    & -     & -     & -   \\
\hline

\end{tabular}
\caption{Solution quality (cost) and running time (time) for each
  algorithm to minimize the function F7 with a minimum value of
  $0$. The algorithms are ranked from the best to the worst on the
  average solution quality.}
\label{tab:f7}
\end{table}

\begin{table}[ht]
\centering
\begin{tabular}{r|ccccccc|cc}

\hline
\multicolumn{10}{c}{{\bf Solution quality ranks on each function}}\\
\hline\hline
\multicolumn{1}{c}{} &
\multicolumn{7}{|c}{Functions} &
\multicolumn{2}{|c}{Average}\\
\multicolumn{1}{c|}{Algorithm} &
\multicolumn{1}{c}{F1} &
\multicolumn{1}{c}{F2} &
\multicolumn{1}{c}{F3} &
\multicolumn{1}{c}{F4} &
\multicolumn{1}{c}{F5} &
\multicolumn{1}{c}{F6} &
\multicolumn{1}{c}{F7} &
\multicolumn{1}{|c}{Arithmetic} &
\multicolumn{1}{c}{Geometric}\\
\hline
KLS2 & 1 & 1 & 4 & 2 & 2 & 3 & 1 & 2.00 & 1.74 \\
SA   & 1 & 4 & 2 & 3 & 6 & 1 & 2 & 2.71 & 2.25 \\
GA   & 1 & 2 & 1 & 5 & 1 & 6 & 6 & 3.14 & 2.34 \\
KLS  & 1 & 3 & 3 & 4 & 3 & 2 & 3 & 2.71 & 2.52 \\
KLS1 & 1 & 6 & 6 & 1 & 4 & 5 & 5 & 4.00 & 3.22 \\
Hill & 1 & 5 & 5 & 6 & 5 & 4 & 4 & 4.29 & 3.83 \\
\hline

\end{tabular}
\caption{The rank of each algorithm on each function F1 to F7. The
  last columns show the average arithmetic and geometric ranks for
  each algorithm where lower rank is better. The algorithms are ranked
  in ascending order of the average geometric rank.}
\label{tab:ranks}
\end{table}

\section{Results and Discussion}
\label{sec:results}

The test results are summarized in Table~\ref{tab:f1f2} for F1 and F2,
Table~\ref{tab:f3f4} for F3 and F4, Table~\ref{tab:f5f6} for F5 and
F6, and Table~\ref{tab:f7} for F7. For the solution quality (Cost in
the tables), these tables give the average (Avr) cost, the standard
deviation (Std) of the costs, the minimum (Min) cost, and the maximum
(Max) cost in 50 runs; and for the running time (Time in the tables),
these tables give the average and standard deviation of the running
times in 50 runs. In each table, the algorithms are ranked from the
best to the worst in terms of their solution qualities.

The results for simulated annealing (SA) and genetic algorithm (GA) in
Table~\ref{tab:f5f6} and Table~\ref{tab:f7} contain ``-''s since the
running times were very large in those cases. For example, I had to
kill the process for the SA algorithm on F5 after about 22 hours. For
ranking of those runs with any results, I assumed that the solution
quality was also the worst.

A value of $0.00$ for running time means that the running time is less
than one second. All the figures are given up to two significant
digits after the decimal point to facilitate their
interpretation. Also, the standard deviation figures for the genetic
algorithm could not be obtained, hence, ``-'' in these tables, since
the GAucsd package does all the runs together.

The running times of the SA and GA algorithms are usually the
largest. The running times of the hill climbing algorithm is the
smallest as expected since it employs the smallest number of
moves. The running times of the KLS, KLS1, and KLS2 algorithms are in
between. The running time of the KLS algorithm like the SA algorithm
on F5 is very large since the landscape of F5 contains many close
hills and valleys, making it harder for local search algorithms. Note
that as expected, the running times of the KLS1 and KLS2 algorithms
are much smaller than that of the KLS algorithm since they explore
smaller search spaces.

As for the solution quality, refer to Table~\ref{tab:ranks}, which
shows the solution quality rank of each algorithm on each function as
well as the average arithmetic and geometric rank of each algorithm
over all functions. KLS2 has the top rank on both average rankings
whereas KLS is the second best on the average arithmetic ranking and
the fourth best on the average geometric ranking. SA and GA are in the
middle with KLS1 and hill climbing at the bottom ranks.

Overall, excluding KLS2's problem structure advantage, if the
algorithms are grouped into tiers with respect to the solution
quality, it seems SA followed by GA are the top tier; KLS is in the
next tier, and hill climbing is in the bottom tier; with respect to
the running times, the tier ranking is almost the reverse. All in all,
the performance of KLS confirms the premise of the variable depth
search approach, as predicted by Kernighan and Lin; moreover, the
approach leads to algorithms (like KLS, KLS1, KLS2) that provide a
great tradeoff between the solution quality and the running time.

In addition to these general observations, as the performance of KLS2
shows, taking advantage of the problem structure can lead to
algorithms that can beat even the top tier algorithms. At the same
time, as the performance of KLS1 shows, this is not always true. What
matters is how the problem structure is taken advantage of. Learning
from these, it seems advisable to try different knowledge
representations and move definitions when a problem structure offers
clues to specialization.

\section{Conclusions}
\label{sec:conclusions}

In this paper, a generalization of the variable depth search mechanism
of the Kernighan-Lin and Lin-Kernighan algorithms, called the KLS
algorithm, is presented. The KLS algorithm was evaluated in comparison
with the hill climbing algorithm, the simulated annealing algorithm,
and the genetic algorithm for function optimization on a standard test
suite consisting of seven functions. The experimental results indicate
that the KLS algorithm is a good candidate for combinatorial
optimization based on the solution quality and the running time
combined.

Future work on the KLS algorithm may include the application of the
algorithm to various combinatorial optimization problems and the
design of its new versions that incorporate more knowledge-intensive
approaches in the search process. Since the problem of learning while
searching is very difficult~\cite{Ac89}, the KLS algorithm may suffer
from the same problems that the other local search algorithms suffer.

\section*{Acknowledgments}

I thank Dr. Yakup Genc for generating the three- and two-dimensional
plots of the functions in my test suite. I thank Professor DeJong for
his comments and grading on my class project report. Finally, I thank
the authors and maintainers of the open-source Tesseract and
Pytesseract software packages for optical character recognition. I
used these packages to convert most of the text of this paper from the
scanned copy of the 1995 class report.

\bibliographystyle{plain}
\bibliography{kl}

\end{document}